\documentclass{desyproc}

\usepackage{amsmath}
\usepackage{amssymb}
\usepackage{esint}

\begin{document}
\contribID{familyname\_firstname}

\desyproc{DESY-PROC-2012-04}
\acronym{Patras 2012}
\doi

\title{Isocurvature Constraints and Gravitational Ward Identity}

\author{Hojin Yoo\\
[1ex]Physics Department, University of Wisconsin-Madison, Madison,
WI 53706}
\maketitle
\begin{abstract}
Axions and WIMPZILLAs are well-motivated dark matter candidates with
interesting cosmological implications, such as isocurvature perturbations
and non-Gaussianities. However, these predicted implications in the
literature are based on the assumption that the cross-correlation
between curvature and CDM isocurvature is negligible. This contribution
discusses the cross-correlation in the axion and the WIMPZILLA scenarios.
Particularly, it is shown that the gravitational Ward identity associated
with diffeomorphism invariance plays an important role in this cross-correlation
calculation confirming the assumption.
\end{abstract}

\section{Introduction}

Currently, the cosmological observational data provides strong support
for the simplest single field slow-roll inflationary models predicting
nearly scale-invariant, adiabatic and Gaussian density perturbations.
However, there still exist many inflationary models compatible with
the current data. These models may have interesting deviations from
the prediction of the simplest inflationary models. In particular,
the axion and the WIMPZILLA scenarios \cite{Seckel:1985tj,Chung:2011xd}
in inflationary models can give rise to CDM isocurvature perturbation
and non-Gaussianities, which are detectable by ongoing and near future
experiments.

In the literature, the cosmological implication of these scenarios
has been studied with the assumption that the cross-correlation between
curvature perturbation and isocurvature perturbation is negligible.
The careful prediction of the cross-correlation is very important
because the current observational constraints of the CDM isocurvature
significantly depend on the cross-correlation. However, the validity
of the assumption is not evident because of the gravitational interaction
of perturbations.

In this contribution, we discuss the curvature and isocurvature cross-correlation
$\left\langle \zeta\delta_{S}\right\rangle $ for the axion and the
WIMPZILLA scenarios. Particularly, we emphasize the importance of
diffeomorphism invariance in the computation of the cross-correlation.
We present the result of the computation, which shows that the cross-correlation
is too small to be detectable through the current CMB experiments,
and this provides proof of the validity of the assumption.

\section{Observational constraints on isocurvature perturbations}

The current observational data shows that the CMB power spectrum is
consistent with the adiabatic initial condition. However, it does
not completely rule out contributions from CDM isocurvature, but it
provides the constraints on the primordial isocurvature perturbations,
which is usually parameterized by two variables
\[
\alpha\equiv\frac{\Delta_{\delta_{S}}^{2}}{\Delta_{\zeta}^{2}+\Delta_{\delta_{S}}^{2}},\quad\beta\equiv\frac{\Delta_{\zeta\delta_{S}}^{2}}{\sqrt{\Delta_{\zeta}^{2}\Delta_{\delta_{S}}^{2}}},
\]
where $\Delta_{\zeta}^{2},\,\Delta_{\delta_{S}}^{2}$, and $\Delta_{\zeta\delta_{S}}^{2}$
are the power spectra of adiabatic, isocurvature perturbations, and
their cross-correlation, respectively, at the primordial epoch. The
isocurvature contribution to the CMB temperature perturbation is expected
to be roughly less than 10\% compared to the curvature contribution.
More precisely, based on the WMAP+BAO+$H_{0}$ data \cite{Komatsu:2010fb},
\[
\alpha_{0}<0.077\mbox{ (95\% CL), and }\alpha_{-1}<0.0047\mbox{ (95\% CL)},
\]
where the subscript denotes the parameter $\beta$, which is the fractional
cross-correlation. The significant difference in the upper-bound of
$\alpha$ between uncorrelated and totally (anti-)correlated cases
originates from the different behaviors of the adiabatic and isocurvature
radiation transfer functions. In particular, the transfer function
of isocurvature perturbations on small scales is suppressed by an
additional factor $k_{eq}/k$ compared to that of adiabatic perturbations.
Therefore, when the cross-correlation is not negligible $\Delta_{\zeta\delta_{S}}^{2}\sim\sqrt{\Delta_{\zeta}^{2}\Delta_{\delta_{S}}^{2}}$,
the cross-correlation contribution $\Delta_{\zeta\delta_{S}}^{2}$
to the CMB temperature perturbation is generally larger than the pure
isocurvature contribution $\Delta_{\delta_{S}}^{2}$ on small scales.
This explains why the upper limit of $\alpha$ for totally (anti-)correlated
models is tighter than that for uncorrelated models.

Therefore, estimating the cross-correlation is crucial to give correct
predictions and restrict the parameters of isocurvature models. For
instance, the axion scenario with a negligible homogeneous vacuum
misalignment angle (and similarly the WIMPZILLA scenario with a negligible
homogenous background field value) predicts detectable non-Gaussianity
\cite{Chung:2011xd,Hikage:2008sk}
\[
f_{NL}\sim30\left(\frac{\alpha}{0.067}\right)^{3/2}
\]
if the assumption that the cross-correlation is zero or negligible
is valid. In the following section, we discuss the cross-correlation
for the axion and the WIMPZILLA scenarios.

\section{Cross-correlation for axion and WIMPZILLA scenarios}

We consider the axion and the WIMPZILLA scenarios discussed in Refs.
\cite{Chung:2011xd,Fox:2004kb}. Particularly, we assume that Pecci-Quinn
symmetry is spontaneously broken during inflation for axions. In both
scenarios, when the spatial average value of the axion or the WIMPZILLA
is much less than its inhomogeneity, the isocurvature perturbation
generated from their density perturbations in the comoving gauge is
written as
\[
\delta_{S}=\omega\frac{\sigma^{2}-\left\langle \sigma^{2}\right\rangle }{\left\langle \sigma^{2}\right\rangle },
\]
where $\sigma$ denotes an axion or WIMPZILLA field, and $\omega\equiv\Omega_{\sigma}/\Omega_{CDM}$.
For axions, $\sigma=f_{a}\delta\theta_{a},$ where $f_{a}$ is the
PQ symmetry breaking scale and $\delta\theta_{a}$ is the inhomogeneity
of the phase of axion.

In order to obtain the cross-correlation $\left\langle \zeta\delta_{S}\right\rangle $,
we calculate the correlator $\left\langle \zeta\sigma^{2}\right\rangle $
using ``in-in'' formalism \cite{Weinberg:2005vy}. We only consider
the gravitational coupling whose interaction Hamiltonian is derived
from the ADM formalism with a gauge choice. With the comoving gauge$\ensuremath{(\delta\phi=0)}$,
the cubic interaction Hamiltonian is
\begin{eqnarray*}
H_{\zeta\sigma\sigma}^{I}(t) & = & -\frac{1}{2}\int d^{3}x\, a^{3}(t)\, T_{\sigma}^{\mu\nu}(t,\vec{x})\delta g_{\mu\nu}(t,\vec{x}),\\
\delta g_{\mu\nu} & = & \left(\begin{array}{cc}
-2\frac{\dot{\zeta}}{H} & \partial_{i}\left(-\frac{\zeta}{H}+\epsilon\frac{a^{2}}{\nabla^{2}}\dot{\zeta}\right)\\
\partial_{i}\left(-\frac{\zeta}{H}+\epsilon\frac{a^{2}}{\nabla^{2}}\dot{\zeta}\right) & a^{2}\delta_{ij}2\zeta
\end{array}\right),
\end{eqnarray*}
where $T_{\sigma}^{\mu\nu}$ is the stress energy tensor of the field
$\sigma$, and $\delta g_{\mu\nu}$ is the scalar metric perturbation.
Then the two-point correlator is written as
\begin{equation}
\widetilde{\left\langle \zeta\sigma^{2}\right\rangle _{p}}=\int d^{3}x\, e^{-i\vec{p}\cdot\vec{x}}\int^{t}d^{4}z\, a^{3}(t_{z})\left\langle \left[\zeta(t,\vec{x})\sigma^{2}(t,0),\frac{i}{2}\left(2\zeta a^{2}\delta_{ij}T_{\sigma}^{ij}\right)_{z}\right]\right\rangle +\mathcal{O}\left(\frac{p^{2}}{a^{2}}\right),\label{eq:sigma2zeta}
\end{equation}
where the contributions from other interaction terms are $\mathcal{O}\left(p^{2}/a^{2}\right)$
because they are derivatively coupled with $\zeta$. One might try
to estimate the integral using the super-horizon approximation frequently
used to compute correlators in the inflationary de Sitter background,
which yields a large cross-correlation. However, the estimation fails
because of the explicit breaking of diffeomorphism invariance.

In fact, the integral (\ref{eq:sigma2zeta}) turns out to be small
because of the Ward identity associated with diffeomorphism invariance,
especially the spatial dilatation, i.e., $x^{\mu}\to\tilde{x}^{\mu}+\lambda X^{\mu},$
where $X^{\mu}(x)=(0,x^{1},x^{2},x^{3}).$ Notice that the curvature
perturbation $\zeta$ is smoothly extended to the spatial dilatation
transform by taking the external momentum $\vec{p}$ to zero \cite{Weinberg:2003sw}.
This fact with the Ward identity allows us to factorize $\left\langle \zeta\zeta\right\rangle $
from a correlator involving external $\zeta$ in the $p\to0$ limit.
The well-known example is the consistency relation for the 3-point
function of $\zeta$ in the squeezed limit \cite{Maldacena:2002vr}
\[
\left\langle
  \zeta_{\vec{p}_{1}}\zeta_{\vec{p}_{2}}\zeta_{\vec{p}_{3}}\right\rangle
\overset{p_{1}\to0}{\longrightarrow}-(2\pi)^{3}\delta^{3}(\sum_{i}\vec{p}_{i})\left|\zeta_{p_{1}}^{o}\right|^{2}\left|\zeta_{p_{2}}^{o}\right|^{2}\frac{\partial}{\partial\ln
  p_{2}}\ln p_{2}^{3}\left| \zeta_{p_{2}}^{o}\right|^{2},
\]
where $\langle\zeta_{\vec{p}_{1}}\zeta_{\vec{p}_{2}}\rangle=(2\pi)^{3}\delta^{3}(\vec{p}_{1}+\vec{p}_{2})\left|\zeta_{p_{1}}^{o}\right|^{2}$.
Similarly, in the two-point function case, we find
\[
\widetilde{\left\langle \zeta\sigma^{2}\right\rangle _{p}}\overset{p\to0}{\longrightarrow}\left|\zeta_{p}^{o}\right|^{2}\frac{\partial}{\partial\ln a}\left\langle \sigma^{2}\right\rangle .
\]
Note that here we assume that UV divergences are properly treated.
See Ref. \cite{crosscorr-inprep} for details.

In our recent work \cite{crosscorr-inprep}, this computation has
been done explicitly, and we have found that
\begin{eqnarray}
\widetilde{\left\langle \zeta\sigma^{2}\right\rangle _{p}} & = & \left|\zeta_{p}^{o}\right|^{2}\times\begin{cases}
\frac{\Gamma^{2}(\nu)H^{2}}{\pi^{3}}\left(\frac{p}{2aH}\right)^{3-2\nu} & \mbox{for scalar in de Sitter space}\\
\frac{H_{p}^{2}}{4\pi^{2}} & \mbox{for massless scalar}
\end{cases}+O\left(\frac{p^{2}}{a^{2}}\right),\label{eq:sigma2zeta_ds_CG}
\end{eqnarray}
where $H_{p}$ denotes the Hubble scale at which scale p exits the
horizon, $\nu\equiv\sqrt{9/4-m_{\sigma}^{2}/H^{2}}$ and $p^{3}\left|\zeta_{p}^{o}\right|^{2}=2\pi^{2}\Delta_{\zeta}^{2}=H_{p}^{2}/4M_{p}^{2}\epsilon$.$ $
In addition, in the dS space-time, we have
\begin{eqnarray}
\widetilde{\left\langle \sigma^{2}\sigma^{2}\right\rangle _{p}} & = & \frac{1}{2\pi^{2}}\frac{H^{4}}{p^{3}}\frac{1}{3-2\nu}\left(\frac{p}{aH}\right)^{6-4\nu}\left[1-\left(\frac{\Lambda_{IR}}{p}\right)^{3-2\nu}\right],\label{eq:sigma2sigma2}
\end{eqnarray}
where $m_{\sigma}<3H/2$. Combining the results (\ref{eq:sigma2zeta_ds_CG})
and (\ref{eq:sigma2sigma2}) yields the fractional cross-correlation
$\beta=\widetilde{\left\langle \zeta\sigma^{2}\right\rangle }/\sqrt{\widetilde{\left\langle \sigma^{2}\sigma^{2}\right\rangle }\left|\zeta_{p}^{o}\right|^{2}}\lesssim\frac{\Delta_{\zeta}}{2},$
which is quite small; however, this result is still interesting because
the cross-correlation is not decaying, and because it is independent
of the parameters, such as $m_{\sigma}$ and $\omega$.

\section{Discussion and Conclusion}

The above result shows that the fractional cross-correlation $\beta\lesssim\Delta_{\zeta}/2\approx2.5\times10^{-5}$
in both scenarios, which is too small to be measured through the current
CMB experiment. More precisely, in order to have approximately the
same level of the CMB power spectrums from pure isocurvature and cross-correlation
at the intermediate scale $l\sim200$, i.e. $C_{l}^{iso}\sim C_{l}^{cor}$,
the fractional cross-correlation should be at least $\beta\sim4\times10^{-2}$.

The smallness of the cross-correlation is understood by the fact that
the super-horizon mode of the curvature perturbation $\zeta$ can
be smoothly extended to the gauge mode, which is the spatial dilatation
in the $p\to0$ limit. Physically, the curvature perturbation $\zeta$
can affect the particle density $\rho_{\sigma}$ and generate small
cross-correlation only at its horizon crossing, since $\zeta$ freezes
after horizon exit and effectively becomes a gauge mode. Thus this
verifies the assumption that the cross-correlation is negligible in
the axion and the WIMPZILLA scenarios.

\section{Acknowledgements}

This work has been done with Daniel J. H. Chung and Peng Zhou. HY
would like to thank the organizers of the 8th Patras Workshop for
financial assistance and the opportunity to present this work.

\begin{footnotesize}

\end{footnotesize}
\end{document}